# Machine learning assisted optical diagnostics on a cylindrical surface dielectric barrier discharge

D. Stefas[1,+], K. Giotis,[1,2+] L. Invernizzi[1], H. Höft[3], K. Hassouni[1], S. Prasanna[1], P. Svarnas[2,*], G. Lombardi[1], and K. Gazeli[1,*]

[1]*Laboratoire des Sciences des Procédés et des Matériaux (LSPM—CNRS), Université Sorbonne Paris Nord, Villetaneuse F-93430, France*
[2]*High Voltage Laboratory (Plasma Technology Room), Department of Electrical and Computer Engineering, University of Patras, Rion Patras 26504, Greece*
[3]*Leibniz Institute for Plasma Science and Technology (INP), Felix-Hausdorff-Str. 2, 17489 Greifswald, Germany*

[*]kristaq.gazeli@lspm.cnrs.fr and svarnas@ece.upatras.gr
[+]these authors contributed equally to this work

**ABSTRACT**

The present study explores combining machine learning (ML) algorithms with standard optical diagnostics (such as time–integrated emission spectroscopy and imaging) to accurately predict operating conditions and assess the emission uniformity of a cylindrical surface Dielectric Barrier Discharge (SDBD). It is demonstrated that these optical diagnostics can provide the input data for ML which identifies peculiarities associated with the discharge emission pattern at different high voltage waveforms (AC and pulsed) and amplitudes. By employing unsupervised (Principal Component Analysis (PCA)) and supervised (Multilayer Perceptron (MLP) neural networks) algorithms, the applied voltage waveform and amplitude are categorised and predicted based on correlations/differences identified within large amounts of corresponding data. PCA allowed us to effectively classify the voltage waveforms and amplitudes applied to the SDBD through a transformation of the spectroscopic/imaging data into principal components (PCs) and their projection to a two-dimensional PC space. Furthermore, an accurate prediction of the voltage amplitude is achieved using the MLP which is trained with PCA–preprocessed data. A particularly interesting aspect of this concept involves examining the uniformity of the emission pattern of the discharge. This was achieved by analysing spectroscopic data recorded at four different regions around the SDBD surface using the two ML–based techniques. These discoveries are instrumental in enhancing plasma–induced processes. They open up new avenues for real–time control, monitoring, and optimization of plasma–based applications across diverse fields such as flow control for the present SDBD.

## I. INTRODUCTION

For a long time now, the experimental characterisation of atmospheric pressure plasmas (APPs) has been a challenging task because of their transient nature, deviation from thermal equilibrium, relatively small dimensions, etc. To achieve a profound description of key plasma quantities, advanced diagnostics allowing for localized and time–resolved measurements should be employed. Indicative examples refer to laser induced fluorescence for absolute density and quenching coefficients of reactive species, absorption spectroscopy for densities and effective lifetimes of metastable states, and Intensified Charge Coupled Device (ICCD) cameras for discharge dynamics [1–8]. Despite their strong capacity, these diagnostics often require sophisticated/expensive equipment arrangements and cumbersome data post–processing methodologies which elevate their cost, limit their portability, and increase their complexity, thus impeding their routine use in practical cases where fast and real–time monitoring of basic plasma properties is required. Thus, it would be highly desirable to employ more compact experimental tools and direct data–driven methodologies in order to achieve a complementary description of global plasma features. For instance, to gain basic knowledge about the origin and morphology of emission patterns in APPs as well as their electrical, thermal, and physicochemical properties, electrical measurements using voltage/current probes, surface temperature monitoring via infrared cameras, space– and time–integrated optical emission spectroscopy (OES) using compact/standardised spectrometers, as well as imaging using CCD sensors, have been employed [9–12]. These are easy to implement, non–invasive and allow for real–time characterisation of the plasma. In particular, they can provide information about the overall discharge morphology, global gas temperature, electron properties and/or relative electric field, among other plasma characteristics. Furthermore, they are compatible with powerful data–driven methodologies such as Machine Learning (ML), which offer a better visualisation, classification, and interpretation of large amounts of raw experimental data [13–16]. This coupled approach may enable an unprecedented classification and reveal otherwise unseen correlations between various experimental datasets recorded, thus leading to an improved understanding and control of the plasma behavior.

The integration of ML practices into the field of plasma physics and related technologies (fusion, biomedicine, material synthesis, agriculture) has marked notable progress [13–23]. ML is a powerful tool for assisting plasma experiments and models since it can be informed from relevant datasets and achieve automated and predictive control of plasma parameters. Typical data-driven ML architectures encompass multivariate statistics and unsupervised/supervised learning algorithms [14]. These allow better monitoring of global plasma quantities, detecting complex patterns between large datasets, clustering and correlating unlabelled data, etc. Principal Component Analysis (PCA) is one of the most commonly employed unsupervised architectures. PCA can be directly informed from standard diagnostics (e.g., OES, CCD, electrical), and reveal information about the behaviour of a plasma source which is not accessible with these diagnostics alone [20]. Besides, it has shown promise in simplifying plasma kinetic simulations, e.g., by reducing the size of reaction sets used in global models of non–equilibrium plasmas [13,19]. In addition, Artificial Neural Networks (ANNs) represent a typical supervised architecture which can understand more complicated data relationships thanks to having one or more hidden layers. This makes ANNs complementary with plasma diagnostics and models, thus forming a "virtual instrument" which can predict plasma instabilities and reveal interrelations between fundamental quantities (electron density $n_e$ and temperature $T_e$, gas temperature, reaction kinetics, etc.) [18]. Therefore, coupling plasma diagnostics and simulations with different ML architectures may enable an extended insight to previously unknown plasma peculiarities and facilitate process control for various applications.

Up to now, most of the relevant literature has focused on the implementation of ML models to low–pressure or dense plasmas [24–30]. Besides, ML–driven analyses of emission spectra have shown new routes for controlling plasma–induced processes like etching and compound detection/classification [28–31]. In most cases, ML models are usually combined with standard electrical and optical methods and extract more details about electron properties ($n_e$ and $T_e$) as well as rotational and vibrational temperatures of probe molecules. For instance, Nishijima *et al.* [24] recorded OES data from a deuterium plasma and measured intensity ratios from three Balmer lines ($D\alpha/D\gamma$ and $D\alpha/D\beta$). The experimental data was used to inform a supervised ML algorithm for the prediction of two–dimensional profiles of $n_e$ and $T_e$. The results obtained showed good agreement with measurements performed with a Langmuir probe. The inclusion to the input data of the intensity ratio of $D_2$ band and $D\alpha$ line improved the prediction efficiency of the algorithm. As another example, Datta *et al.* [25] analysed experimental emission spectra from a high energy density plasma. They employed different supervised ML algorithms (e.g., linear regressor, k-nearest neighbour, decision trees, and random forest) which were trained using synthetic OES data and the outputs were compared to the experimental spectra recorded. These ML models were coupled with OES and their relative performance was assessed for improving the prediction efficiency of $T_e$ and ion density.

In what concerns APPs and their applications, the consideration of ML methods has recently drawn significant attention [14–16, 20–22, 31–35]. In fact, the idea of applying routine diagnostics to APPs that allow probing their global properties and feeding them to powerful ML for subsequent (or real–time) analysis may lead to improved descriptions, similarly to the case of low–pressure plasmas. For instance, Lazarus *et al.* [32] and Bong *et al.* [33] combined distinct ML models with conventional images (captured with inexpensive CCD detectors) of two different APPs. These coupled approaches made it possible to infer the colour and shape of a plasma effluent and achieve real–time determination of distinct plasma features such as propagation length, process gas concentration and flow rate. As another example, Shah Mansouri *et al.* combined OES spectra of an atmospheric pressure plasma jet with a classification algorithm (partial least squares–discriminant analysis) and managed to detect trace $CH_4$ amounts as low as 1 ppm in the plasma [34]. Finally, Lin *et al.* showed that it could be even feasible to develop ANNs which can be informed by physics laws and combine them with quantitative diagnostics such as infrared absorption spectroscopy and predict global plasma quantities such as $n_e$, $T_e$, gas temperature, as well as concentrations of key reactive species (such as $O_3$, $N_2O$, and $NO_2$) [35].

A distinct category of APPs where ML can provide unprecedented insights refers to the surface dielectric barrier discharges (SDBDs) [36–38]. SDBDs are strongly non–equilibrium discharges which have drawn particular attention for applications in flow control, medical and food treatments [36, 37, 39–41]. These are commonly generated in ambient air by means of sinusoidal (AC) and pulsed high voltages of audio frequencies. A typical SDBD arrangement consists of a flat dielectric surface which is always in contact with two electrodes with one of them being the powered electrode and the other being connected to the ground. To achieve a profound characterisation of SDBDs, the implementation of advanced diagnostics such as laser Doppler and particle image velocimetry, Schlieren and streak camera imaging, and picosecond electric field second harmonic generation, among others, would be ideally preferable [38, 42–44]. However, in more realistic cases relatively simpler diagnostics such as voltage/current probes, OES, and ordinary CCD (or ICCD) imaging are accessible. Despite their rather limited performance versus the more advanced diagnostics, they still allow for real–time monitoring of global SDBD features (deposited power, electron properties, plasma temperatures, relative electric field, discharge morphology/dynamics) [38, 41, 45–50]. At the same time, since they can be performed using compact and flexible equipment, they offer the possibility for direct coupling with powerful ML architectures [16]. This approach is advantageous over traditional/manual data treatment methods that often fail to classify, analyse, and diversify hidden information in large amounts of raw data. Feeding such data groups to ML makes it possible to smartly digest them and reveal major directions of variations due to changes in the SDBD properties. To date, only a few works have used this coupled approach referring to linear SDBDs. These mainly focus on improved predictions of SDBD regimes (diffuse/filamentary), induced flow velocities/instabilities and consumed power, autonomous enhancement of plasma effects on vortex formation of separated shear layers, actuator burst frequency optimisation for closed–loop flow separation control over airfoils, and decrease of drag and lift forces for improving flow stability around square cylinders [51–57]. Therefore, more studies are needed to demonstrate different aspects of ML incorporation to the field of SDBDs and their applications.

In this work, first the optical features of a cylindrical SDBD arrangement were studied using conventional experimental methods, namely, space– and time–integrated OES and CCD imaging. Then, the results obtained were utilized to inform dedicated complementary ML architectures for directly predicting plasma operating regimes as well as interrogating the uniformity of the discharge emission pattern. Specifically, the shape of the waveform (AC or pulsed) and corresponding amplitude of the operating voltage were categorised and then predicted. This was done by feeding an unsupervised algorithm, Principal Component Analysis (PCA), with numerous raw OES spectra and CCD images. After that, raw spectroscopic data were used to construct a predictive model for the actual voltage amplitude applied to the SDBD, as a proof of concept. To this end, we employed Multilayer Perceptron (MLP), a representative type of supervised ANNs. On top of that, the capacity of this approach in identifying and classifying discharge peculiarities under AC and pulsed operation was interrogated. This was achieved through the assessment of the discharge uniformity by feeding the ML algorithms with raw OES data recorded from well–specified spatial regions around the reactor's surface.

The remainder of the manuscript is organised as follows. Section **IIA** is a description of the SDBD arrangement investigated in this work and the voltage waveforms applied along with indicative corresponding current signals. Section **IIB** provides details on the acquisition methods of the CCD images and OES spectra of the SDBD. Section **IIC** describes the data–driven architectures used for the numerical analysis of the CCD images and OES spectra. Section **III** presents and discusses the corresponding results obtained, and Section **IV** concludes the paper.

## II. EXPERIMENTAL SETUP AND DATA-DRIVEN METHODOLOGIES

### A. SDBD arrangement and electrical excitations.

A schematic representation of the experimental setup is shown in **Fig.1** displaying the discharge arrangement and the applied diagnostics. The SDBD was operated in atmospheric pressure room air (20 °C controlled temperature and 40–50% relative humidity during all the experiments). It consisted of a cylindrical arrangement of two annular stainless–steel electrodes separated by a quartz dielectric tube. The inner electrode (grounded) was fully submerged into transformer oil, with at least 10 times higher dielectric strength compared to air. Thus, the discharge was only formed in the air layer surrounding the SDBD, starting from the edge (100 μm thickness) of the external powered electrode and propagating downward on the outer surface of the quartz. Indicative discharge images are shown at the bottom right of **Fig.1**. These were taken with a digital single lens reflex camera (DSLR; Canon EOS 4000D, 55 mm focal length, F–stop=f/5.6, exposure time=1/30 s, ISO–6400). More details on the plasma source can be found elsewhere [58].

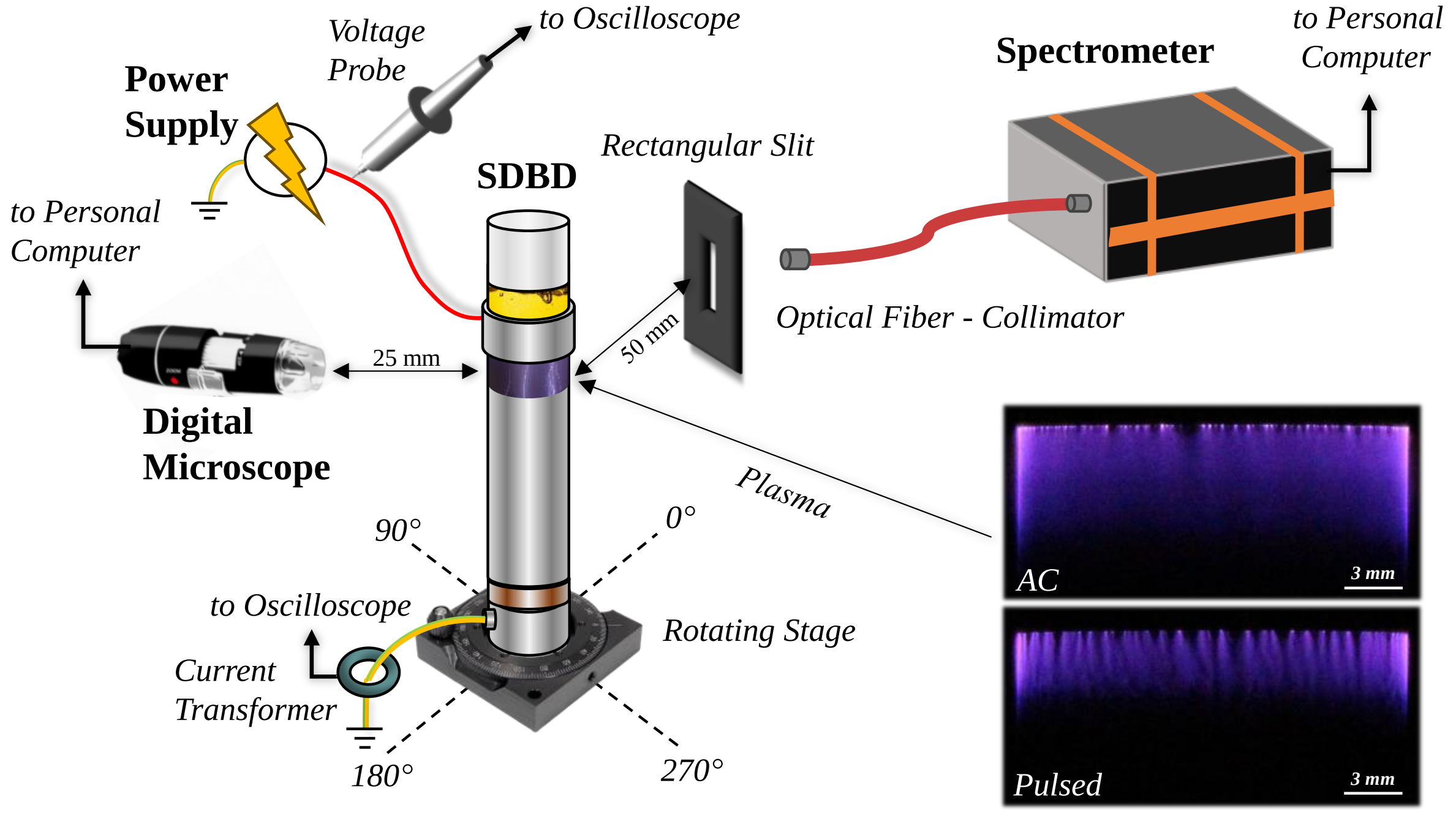

*Fig. 1. Schematic representation of the experimental setup used in this study along with two indicative DSLR SDBD images under sinusoidal AC (20 $kV_{pp}$) and pulsed (10 kV) conditions. The SDBD was placed on a calibrated rotating stage for capturing OES spectra from four different regions/angles with respect to the slit's orientation: 0°(reference position), 90°, 180° and 270°. The use of the slit concerns only the OES studies.*

The voltage waveforms were generated by two high voltage (HV) power supplies: sinusoidal AC (PlasmaHTec®; up to 30 $kV_{pp}$, 10 kHz, 500 W) and unipolar positive pulsed composed of a DC power supply (F.u.G. Elektronik GmbH HCN700–20000), a HV pulse generator (DEI PVX–4110) and a delay/pulse generator (Stanford Research Systems DG645). The voltage and current signals were monitored using a HV probe (Tektronix P6015A; 75 MHz bandwidth, 4–5 ns rise time) and a current transformer (Magnelab CT–D2.5–BNC; 1200 Hz — 500 MHz bandwidth, 0.7 ns rise time) connected to a digital oscilloscope (Teledyne LeCroy WaveSurfer10; 1 GHz bandwidth, 10 GS/s sampling rate). Both HV waveforms exhibited the same repetition frequency (10 kHz), while their amplitudes varied up to 20 $kV_{pp}$ (AC) and 10 kV (pulsed). For the pulsed case, a symmetrical HV pulse waveform was used, i.e., the pulse width was set to 50 μs (equal to the AC half–period).

Representative oscillograms of the two HV waveforms at 20 $kV_{pp}$ (AC) and 10 kV (pulsed) along with corresponding currents are depicted in **Fig.2**. For the AC case (**Fig.2a1,a2**), the number, the peak current and the temporal appearance of the discharge events within a voltage period constantly change over time. For instance, in **Fig.2a1,a2,** numerous erratic current pulses appear during the rising phase of the positive AC half–period (about 25 μs). For the pulsed case, however, only two distinct and stable current pulses are formed (one during the rising and one during the falling phase of the pulse, both with a duration <100 ns) (**Fig.2b1,b2**). Both current waveforms represent the raw total currents recorded with the current transformer which consist of the displacement (capacitive) current due to the capacitive behaviour of the SDBD, and the discharge (drift) current which is superposed on the capacitive current. The corresponding average powers are 5.5 W (AC) and 2.3 W (pulsed).

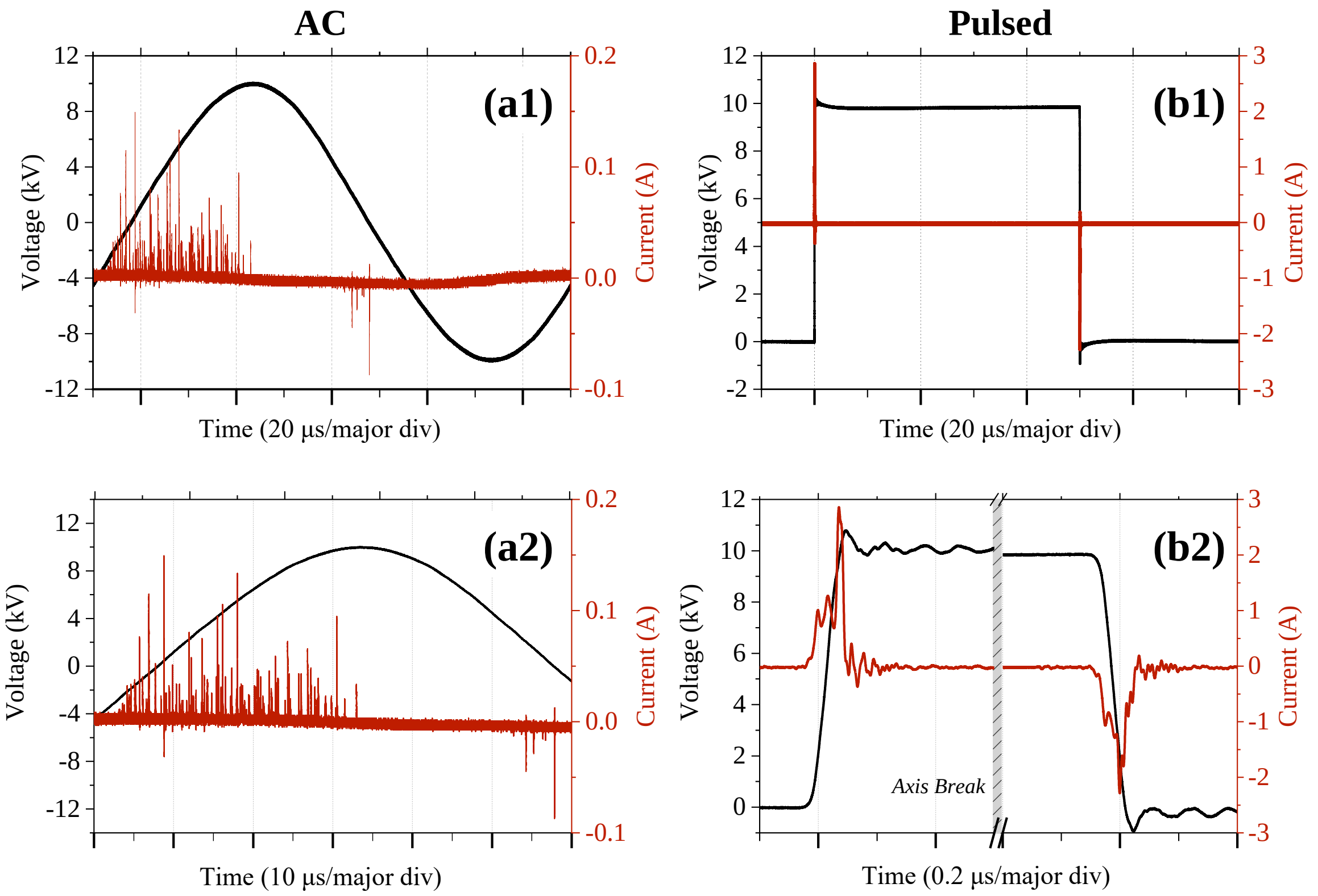

***Fig.2**. (a) Indicative oscillograms of the applied voltages (averaged over 100 signals) and induced total currents (single shots) over one period at 20 $kV_{pp}$ (AC; a1) and 10 kV (pulsed; b1). Figures a2 and b2 depict the voltage–current signals when zooming within the positive half–period of the voltage (AC) and the voltage pulse rising/falling phases (pulsed; 50 ns rise/fall times), respectively; be aware of the break on the time axis in b2.*

The dissimilar behaviour of the SDBD current under AC and pulsed cases is a consequence of the significant differences in the type and the absolute value of $\Delta V/\Delta t$ of the two voltage waveforms (about $4\times10^{8}$ $Vs^{-1}$ for AC, **Fig.2a2**, and about $2\times10^{11}$ $Vs^{-1}$ for pulsed, **Fig.2b2**) which influence the SDBD electrical/optical properties. Xie *et al.* were interested in the effect of the rise time ($r_t$=50–500 ns) of the applied pulsed voltage on the peak power/energy and induced aerodynamic features of a planar SDBD [59]. The highest peak power/energy was obtained for $r_t$=50 ns, reached a minimum for $r_t$=150 ns and then slightly varied up to $r_t$=500 ns. They also recorded Schlieren images of corresponding induced pressure waves (due to the rapid energy release) which had the shape of a circular arc. Similarly to the electrical power, the shock wave corresponding to $r_t$=50 ns was the most prominent in terms of expansion velocity and vertical distance covered. Furthermore, Benard and Moreau investigated the role of AC and ambipolar pulsed voltage waveforms on the electrical characteristics and morphology (among other properties) of a SDBD [48]. Independently of the waveform used, they observed two discharge regimes: (i) filamentary, corresponding to the positive half–cycle with distinct current impulses widely spread in time, and (ii) corona regions during the negative half–cycle which started from the powered electrode and expanded along the dielectric surface. For the AC case, the current peaks formed during the positive (rising) phase of the voltage were noticeably scattered in time (similarly to **Fig.2a1,a2**). However, in the pulsed case, the corresponding current pulses were more condensed in time being superimposed on a large capacitive component. In [48] the rising/falling slopes between the AC and pulsed voltages were slow and comparable between them ($\sim10^{8}$ $Vs^{-1}$; as the AC waveform in **Fig.2a1,a2**), whereas here in this work we used a monopolar positive pulsed voltage with about three orders of magnitude steeper slopes than for the AC case. This results in the formation of two distinct discharge current pulses which are significantly confined in time (duration <100 ns) for the pulsed case (**Fig.2b1,b2**).

**B. CCD imaging and optical emission spectroscopy.**

The discharge emission patterns were captured with a compact portable digital microscope (Jiusion) (**Fig.1**). It was equipped with an RGB CCD detector (640×480 $px^2$ size) with an adjustable focus range from 10 to 250 mm (×40 up to ×1000 magnification, respectively). The microscope operated at a frame rate of 25 FPS (i.e., 40 ms integration time). Since the repetition frequency of the AC and pulsed voltages was set to 10 kHz (i.e., 100 μs period), each CCD image contained discharge light which was integrated over 400 HV periods. For all the analyses performed here, only images with a size of 640×250 $px^2$ were used, thus covering a focused region of 14×5.5 $mm^2$ on the quartz's surface, i.e., mainly the area where the discharges occur.

The spectrally–resolved (200–900 nm) plasma emission was collected using a collimator (COL–UV/VIS; Avantes) mounted in front of the one end of a quartz optical fibre (FC–UV400–2; Avantes). The other end of the fibre was coupled to the entrance slit (50 μm width) of a 75 mm focal length spectrometer (AvaSpec–ULS2048XL; Avantes) (**Fig.1**). The spectrometer was equipped with a 300 lines/mm diffraction grating, and a 2048 px CCD linear image sensor. From this sensor, 423 px were used for ML–based analysis, corresponding to the 280–460 nm spectral range. Between the fibre/collimator and the SDBD, a rectangular slit (5 mm wide) was installed to collect radiation from a SDBD region of 5×5.5 $mm^2$ (**Fig.1**). The position of the slit was invariable while the SDBD was fixed on a calibrated rotating stage which allowed changing its azimuth angle with respect to the slit. In this way, 200 OES spectra (for each voltage amplitude) were recorded at 4 distinct angles around the SDBD: 0° as reference point (coinciding with the slit's orientation in **Fig.1**), 90°, 180° and 270°. Thus, possible artifacts due to spatial overlapping were avoided. Furthermore, attention was paid to maintaining the same 0° position between AC and pulsed cases for reliable analyses. The recorded spectra were integrated in time using 7.5 times larger integration time than the one used for the CCD images (due to a lower signal–to–noise ratio in the case of OES studies) and were utilized to inform the different ML algorithms. The spectroscopic system was calibrated in terms of absolute wavelength using a mercury (UVP 90–0012–01; Analytik Jena) lamp.

### C. ML architectures used for the analysis of CCD images and OES spectra.

Two ML architectures, PCA (unsupervised) and MLP (supervised), were implemented to analyse the experimental data in this work. An essential description of their distinct characteristics is provided in this section, while more details can be found elsewhere [14, 18, 60, 61].

First, PCA allowed us to simplify the complex data recorded by transforming them into a new set of uncorrelated variables through a linear transformation procedure. The procedure followed is schematically represented in **Fig.3a**. The original raw experimental data were stored in matrices of $n \times d$ dimensions, where $n$ was the number of observations (i.e., OES spectra or CCD images) and $d$ was the number of features (i.e., pixels–cells of a matrix containing distinct intensity values for each OES spectrum or CCD image). For the CCD images, the intensity of the SDBD light was distributed along the matrix rows/columns referring to different positions within the recorded SDBD area (14×5.5 $mm^2$; without slit). The light distribution was not homogeneous due to various emissive processes taking place in the recorded discharge area. Similarly, the pixels of the CCD sensor of the spectrometer contained intensity values corresponding to distinct emissive species (wavelengths) emitting within the recorded spectral range. In this case, spatial information was not available since the discharge emission was integrated over the entire area recorded (5×5.5 $mm^2$; with slit). Furthermore, for the CCD images a typical matrix consists of pixels with red (R), green (G), and blue (B) components, each contributing to the overall color and intensity of the image (**Fig.3a**). Therefore, the inputs used for the ML algorithms comprised the intensity values of the pixels of OES spectra or CCD images, which were extracted and utilized for analysis. This allowed us to build the covariance matrix $C$ ($d \times d$ dimensions):

$$C = \begin{pmatrix} cov(x_{1,}x_1) & \cdots & cov(x_{d,}x_1) \\ \vdots & \ddots & \vdots \\ cov(x_{d,}x_1) & \cdots & cov(x_{d,}x_d) \end{pmatrix}$$

where:

$$cov(x, y) = \frac{\sum_{i=1}^{n}(x_i - \bar{x})(y_i - \bar{y})}{n - 1}$$

is the covariance between two features $x$ and $y$ (emission intensity, wavelength, etc.), $x_i$ and $y_i$ are the feature values, $\bar{x}$ and $\bar{y}$ are their mean values. The eigenvectors and eigenvalues of $C$ were computed and sorted according to decreasing eigenvalues using the Singular Value Decomposition (SVD) algorithm [60,62]. The eigenvectors represent the directions of maximum variance and the eigenvalues the magnitude of variance along these directions. In most practical cases, only a small number $k$ of large eigenvalues exists implying the inherent dimensionality of the subspace describing the *"important"* variance of the data, whereas the remaining eigenvalues can be regarded as *"noisy"*. From the aforementioned process, a $d \times k$ matrix, $A$, was formed, whose columns comprised the $k$ eigenvectors. Thus, a representation of the data by principal components (PCs) consisted of projecting the data $u$ onto the $k$–dimensional subspace as:

$$u'_i = A^T(u_i - \mu), i = 1,2,I$$

where μ is the mean vector of the data, $u'_1$ the first eigenvector, i.e., principal component (PC1) which captures the largest amount of variance in the data, $u'_2$ the second (PC2), which captures the second largest amount of variance, and so on (**Fig.3a**). This projection enables data representation in fewer dimensions while still retaining most of the essential information (described through the variance) of the original data [60]. By visually depicting the PCA–transformed data on a scatter plot (where the axes correspond to the PCs), clusters or groups of data points that may be related to each other were identified. The closer two points are to each other, the more similar they are in terms of their overall characteristics. By correlating the eigenvectors of the transformation with the original features, the so–called loadings were extracted. They represent the contribution of each original feature to the respective PCs. A positive loading indicates a positive relationship, while a negative loading indicates a negative relationship. Thus, by simply examining the loadings, we gained insights into the relationship between the data features and the PCs.

Second, MLP is an ANN that comprises an input layer, one or more hidden layers, and an output layer of neurons, each connected to the next layer. It is a fundamental feedforward ANN, widely used for various applications such as pattern recognition, classification, and regression [63]. A schematic representation of its general operating principle is given in **Fig.3b**. The input layer comprises *d* neurons, i.e., the input features of the dataset. In the present work, PCA–preprocessed spectra were used; thus, the input layer comprises *k* neurons. The hidden layers may comprise any number of neurons, and the output layer, in our case, comprises one neuron, where predictions were made (i.e., the voltage amplitude applied to the SDBD). Each neuron received the inputs, performed a computation on them using an activation function, and produced an output that could be passed to other neurons in the network [61]. An activation function determines the output of a neuron given its input. In the present case the hyperbolic tangent, *tanh*, was used. These functions introduce nonlinearity into the network, enabling it to learn complex patterns in data. Furthermore, the connections between neurons are weighted and these weights influence each neuron's output to another neuron's input. During training, the inputs passed through the network many times and the weights were adjusted so that the output of the network matched the desired output, i.e., the neural network predictions ($V_{\text{predicted}}$) matched the actual values ($V_{\text{actual}}$) of the applied HV amplitude [61]. This procedure was performed so that a quantity called "loss function" was minimized, which in the present work was the mean squared error (MSE) of $V_{\text{predicted}}$ against $V_{\text{actual}}$, as follows:

$$\text{MSE} = \frac{1}{n}\sum_{i=1}^{n}(V_{\text{actual}} - V_{\text{predicted}})^2$$

In that concept, a trained MLP model should learn the different patterns of the input data, so that it can accurately predict their voltage amplitude as well as unknown data that haven't been used during the training. The weight optimization was performed using the Adam algorithm by iteratively updating the weights based on the errors the network makes when predicting outputs compared to the actual ones [64].

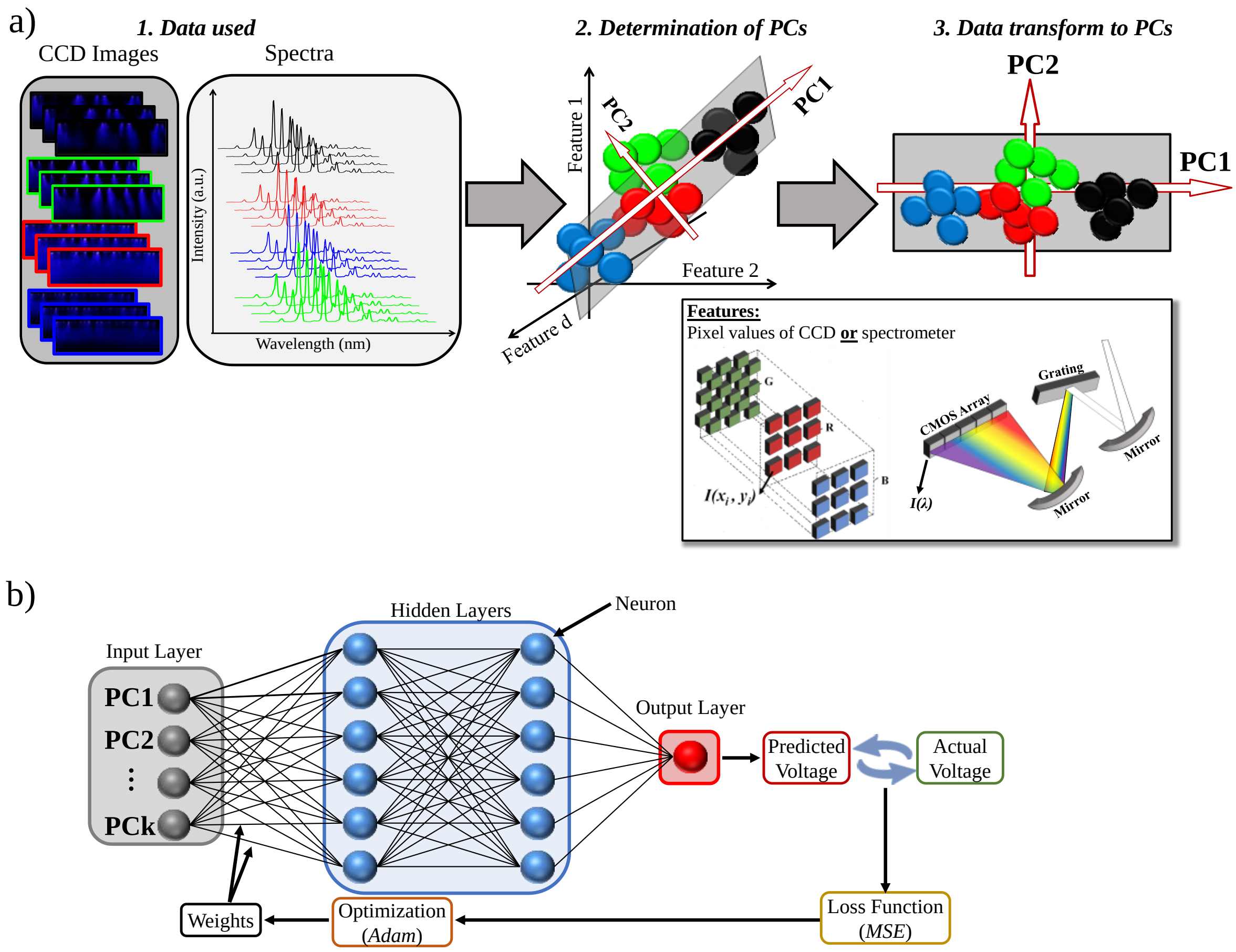

***Fig. 3.*** *Schematic representation of the ML architectures used in this work. a) PCA: CCD images or emission spectra (1.) are projected on a reduced feature space (2.), where the axis represents directions of maximum variance from the original d–dimensional space spanned by the data features (3.). b) MLP: it takes as inputs PCA–pre-processed spectroscopic data. Hidden layers perform non–linear transformations, and the output layer generates voltage amplitude predictions. The weights and the loss function (MSE) optimize the network's performance during training using the Adam algorithm.*

The PCA and MLP algorithms in this work were implemented using the open–source library Scikit–Learn in Python environment [62]. The procedures followed to collect and analyse the CCD and OES data as well as the respective purposes of usage of both ML models are summarised in **Tab.1**.

**Tab.1.** Procedures followed to collect and process various datasets in this work, corresponding ML algorithms used and their purpose. For PCA, the experimental data recorded were directly used for analysis (untrained algorithm), whereas for MLP a data portion was used for its training and the remaining (unknown) data were used for testing its prediction efficiency.

| **Recorded Data (usage)** | **No. of spectra/images** | **ML algorithm employed** | **Purpose of ML usage** | **Comments** |
|---|---|---|---|---|
| • OES at 0° and 90°.<br>With slit (SDBD region recorded: 5×5.5 mm$^2$). | 3600 per angle (200 for each voltage amplitude; 9 AC and 9 pulsed voltage amplitudes) | PCA | Voltage type and amplitude differentiation based on OES pattern visualisation/projection to PCs. Assessment of emission uniformity. | Use of raw (**Fig.4c,d,e**, **Fig.5**) and normalised (**Fig.6**) spectra at 0°. Projection to the 0° PC space of additional spectra (3600) recorded at 90° (**Fig.9**) |
| • OES at 0° (training).<br>• OES at 90°, 180°, and 270° (testing/predicting).<br>With slit (SDBD region recorded: 5×5.5 mm$^2$). | 14400 (3600 for training, 10800 for testing/predicting) | MLP | Actual voltage amplitude prediction and emission uniformity assessment. | Usage of PCA–pre-processed data (**Fig.8**, **Fig.10**) |
| • OES at 0°, 90°, 180°, and 270° (training).<br>• OES at 0° (testing/predicting intermediate voltage amplitudes).<br>With slit (SDBD region recorded: 5×5.5 mm$^2$). | 17600 (14400 for training, 3200 for testing/predicting) | MLP | Actual applied voltage amplitude prediction. | Usage of PCA–preprocessed data (**Fig.11**) |
| • CCD images at 0°.<br>Without slit (SDBD region recorded: 14×5.5 mm$^2$). | 9560 | PCA | Voltage type and amplitude differentiation based on emission pattern visualisation. | **Fig.4a,b**, **Fig.7** |

## III. RESULTS AND DISCUSSION

### A. SDBD emission analyses.

Indicative discharge emission patterns under two representative voltage amplitudes of both waveforms are shown in **Fig.4a,b.** In both cases an increase in the voltage amplitude affects the discharge pattern. For the AC mode (**Fig.4a**) the pattern exhibits numerous strong localized emissions (bright spots) at the powered electrode, which transform to weaker dispersed luminous channels until a vertical distance ($d$) of about 1.5 mm downstream. Then, these channels seem to expand laterally further away from the powered electrode. This results in a somewhat more uniform overall emission profile for $d$>1.5 mm. For the pulsed mode (**Fig.4b**), however, a smaller number of localized strongly emissive spots at the powered electrode is recorded. Then, the discharge evolves within more isolated weaker luminous structures along the entire propagation distance. Thus, discharge bright spots and subsequent propagation channels in the pulsed mode are more localized compared to the AC case. Kettlitz *et al.*

investigated the discharge propagation dynamics in a single filament SDBD arrangement using the same pulsed power supply as in this work [44]. Using an ICCD camera (50 ns gate width) they showed that the bright spots occur most likely during the falling phase of the HV pulse where the powered electrode acts as the cathode.

Actually, the discharge mechanism is not very different between the two operating SDBD conditions, and the peculiarities observed in their emission features may be attributed to the distinct features of the HV waveforms. For AC SDBDs, when the powered electrode is exposed to the positive voltage polarity, it plays the role of the anode. In this regime, also known as the streamer regime, an ionization front is initiated due to electron avalanches accelerated towards the anode and the discharge exhibits different filaments with erratic propagation paths [48,65]. When the powered electrode is negatively biased, it becomes the cathode and a more homogeneous discharge pattern is established [48,65]. For pulsed SDBD of positive polarity, the powered electrode always acts as the anode during the rising phase of the voltage and the discharge is in the streamer regime, depositing positive charges on the dielectric surface [44,49,66]. During the falling voltage phase, initially the powered electrode is at higher potential with respect to the ground, thus still being the anode. However, as the voltage level rapidly decreases to zero (in less than 50 ns), it becomes the cathode as its electric potential becomes lower than that of the dielectric surface which is induced by the accumulated space charge [66]. This will result in a negative current pulse and a more diffuse discharge pattern. The characteristic emission patterns under AC and pulsed modes should be linked with the corresponding discharge currents [48]. These signals are indicatively shown in **Fig.2** referring to a period of the applied voltage. While for the AC case the fingerprints of the conduction current may be directly evident even from the total current in **Fig.2** (i.e., through the formation of positive and negative erratic impulses on it), this distinction is not straightforward in the pulsed case from the same figure. In fact, in this case the $\Delta V/\Delta t$ value is about 3 orders of magnitude larger than the AC case (section **IIA**) and the corresponding capacitive current has a much larger contribution to the total current. Therefore, a subtraction of the capacitive component of the current must be performed for a proper investigation of the discharge current behaviour in both cases. Such a detailed analysis of the electrical signals merits a separate contribution and is not performed in the present work.

In **Fig.4c** representative OES spectra of the SDBD using pulsed voltage are depicted. The main emissive transitions identified lie in the 300–460 nm spectral region and mostly refer to the molecular bands from the Second Positive System (SPS) of $N_2$ and First Negative System (FNS) of $N_2^+$, as follows [67]:

$$N_2(C^3\Pi_u) \rightarrow N_2(B^3\Pi_g) + h\nu \quad (k = 2.74 \times 10^7\ s^{-1})$$

$$N_2^+(B^2X_u^+) \rightarrow N_2^+(X^2\Sigma_g^+) + h\nu \quad (k = 1.6 \times 10^7\ s^{-1})$$

Moreover, additional emissions in the spectral region 600–900 nm were identified (**Fig.4d**), which are representative of the $N_2$ First Positive System (FPS) [67]:

$$N_2(B^3\Pi_g) \rightarrow N_2(A^3\Sigma_u^+) + h\nu \quad (k = 2 \times 10^5\ s^{-1})$$

For a better visualisation of the $N_2$(FPS) spectra, a 65–fold larger integration time was used compared to that in **Fig.4c**. The same emissive species are detected for the AC voltage as well.

For both voltage waveforms, the generation of the excited states of emissive species may be due to reactions taking place during two temporal regimes: (i) active discharge phase (when the discharge current flows), and (ii) post–discharge (afterglow). Besides the radiative relaxations identified under our experimental conditions which lead to the production of $N_2(A^3\Sigma_u^+)$ (metastable) and $N_2(B^3\Pi_g$ ) excited states (**Fig.4**), the upper $N_2$ states $N_2(A^3\Sigma_u^+)$, $N_2(B^3\Pi_g$ , and $N_2(C^3\Pi_u$ ) are mainly populated through electron impacts which are functions of the strength of the reduced electric field (E/N) [67,68].

In addition, a noteworthy result is revealed when comparing OES emission fingerprints between several amplitudes of the two voltage waveforms. Indicatively, **Fig.4e** displays $N_2$(SPS) and $N_2^+$(FNS) bands recorded for two amplitudes of each voltage waveform. First, an enhancement of the species intensities with the increment of the input voltage amplitude in both cases is revealed. Then, the spectral patterns between 7.5 kV (pulsed) and 16 $kV_{pp}$ (AC), and between 10 kV (pulsed) and 20 $kV_{pp}$ (AC) exhibit almost identical intensity profiles. The average electric power in these cases is 1.2 W (pulsed; 7.5 kV), 2.6 W (AC; 16 $kV_{pp}$), 2.3 W (pulsed; 10 kV) and 5.5 W (AC; 20 $kV_{pp}$). The power values of the AC waveform are larger than those of the pulsed case. However, for the AC case, a relatively larger portion of the power is expected to be dissipated into dielectric heating instead of gas excitation and ionization [69]. In **Fig.4e**, without any previous knowledge of the voltage features, identifying different SDBD operating regimes seems impossible by a visual inspection of the OES spectra shown, even for an experienced user. To be able to predict the voltage type, a comparison with corresponding CCD images is necessary which reveals different discharge emission patterns.

However, even when coupling OES spectra with CCD images and manually comparing their corresponding data, a frequent user cannot predict the amplitude of a given voltage waveform applied to the SDBD without previously knowing it. Furthermore, in practical cases such as surface treatments, automated predictions of discharge properties are largely preferred over manual data treatments [9]. This is because they are much faster than frequent users in handling vast datasets and making necessary predictions/adjustments. Therefore, as a proof of concept, in this study we turned our attention to ML architectures for predicting the SDBD operating regimes. Specifically, by informing ML either with OES spectra or CCD images, we were capable of differentiating both the type and the amplitude of the applied voltage to the SDBD. In addition, using ML–driven analyses of the SDBD emission allowed us to assess the discharge uniformity at different well–defined regions around the SDBD arrangement. Therefore, ML can be a strongly supportive tool to both optical diagnostics considered in this work by revealing useful information on various SDBD aspects. The results obtained through this coupling are presented and discussed in Sections **IIB–IIE**.

Achieving accurate predictions of plasma operating conditions is highly appreciable in applications where real–time automated control of global plasma features is demanded. Such an unprecedented performance could be reached through the coupling of standardized electrical and optical diagnostics with dedicated data–driven architectures. In such a scenario, the algorithms developed would receive continuous feedback from experimental data, thus making fast parallel predictions and continuous adjustments to discharge parameters according to the outcomes requested by users. For instance, Gidon *et al.* [9] focused on medical applications of APPs. They used an AC–driven helium plasma jet impinging on a borosilicate microscope cover slip placed on a grounded aluminium plate and studied thermal effects induced on the substrate (measured via an infrared camera). They developed a numerical procedure based on a single–board computer to receive and process experimental data in real–time (gas flow rate, applied peak–to–peak voltage, substrate position and temperature) and make relevant predictions/adjustments of plasma/surface parameters. Using this data–driven methodology, they showed that it is possible to achieve real–time regulation of the delivery of the thermal dose to the substrate. Furthermore, they underlined the benefits of controlled feedback to achieve a uniform thermal dose delivery to the substrate. This approach can be further extended by coupling it with ML–driven data processing methodologies in order to perform real–time diagnosis of more complex plasma parameters such as rotational and vibrational temperatures [16].

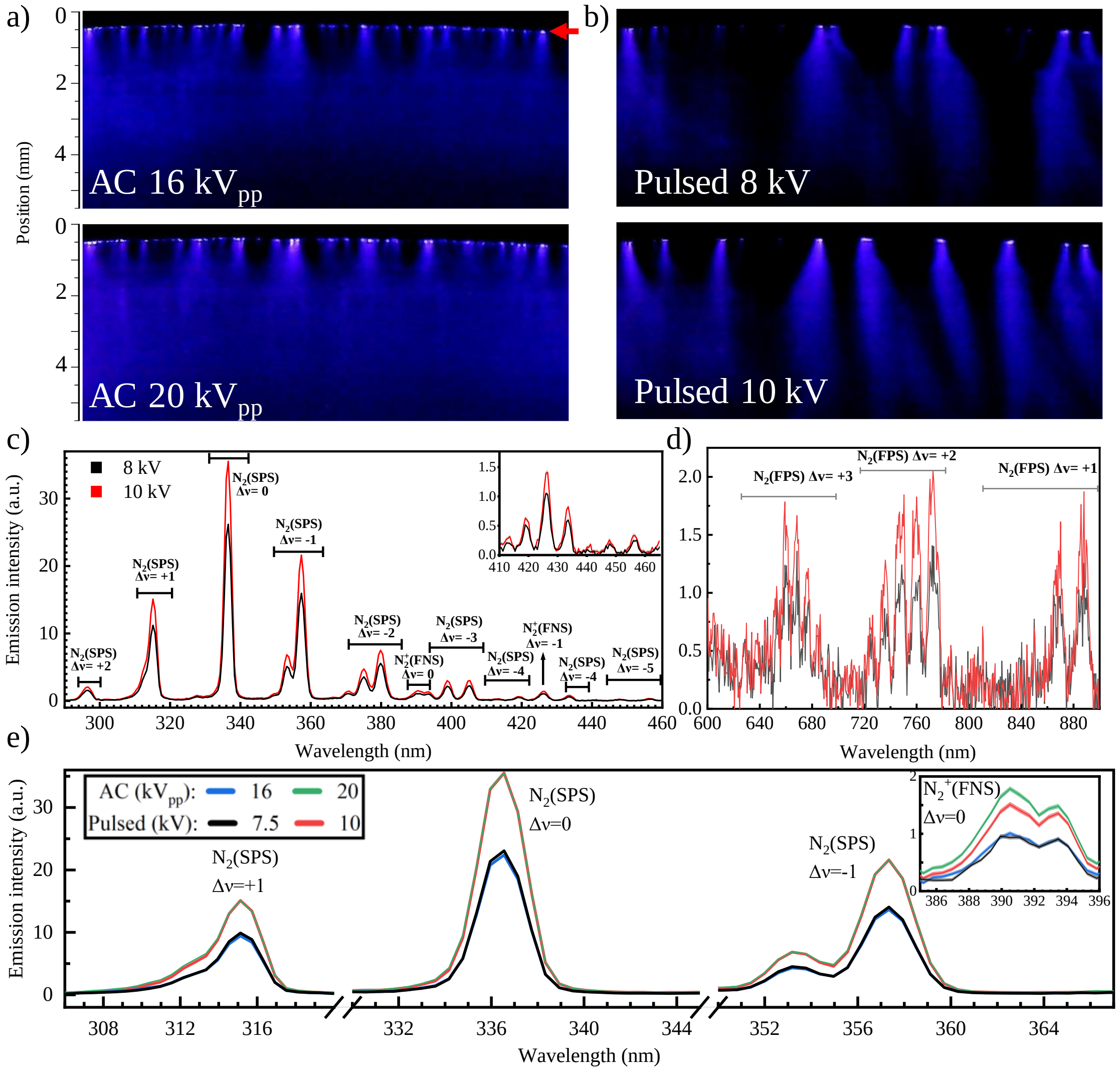

***Fig. 4.*** *CCD images of the SDBD emission pattern using a) AC and b) pulsed voltage; red arrow: edge of the high–voltage electrode. Emission spectra of the SDBD under pulsed voltage in the c) 280–460 nm and d) 600–900 nm spectral ranges (a 65–fold larger integration time was used compared to that in 4c). Bottom: e) comparison of the $N_2$(SPS) and $N_2^+$(FNS) (inset) emission intensities between different types and amplitudes of the applied voltage (same integration time as in 4c). SDBD orientation: 0° angle (**Fig.1**).*

## B. Visualization of patterns in emission spectra and CCD images using PCA.

For a more detailed investigation of the discharge aspects under different operating voltage types and amplitudes, the measurements with the optical microscope and spectrometer were assisted by feeding the PCA with the corresponding data obtained (**Table 1**). First, a reduction of the dimensionality of the collected OES data, and a better visualisation based on their variance was performed. In **Fig.5a** the first two PCs are shown, where the spectra corresponding to AC and pulsed modes are represented with circles and rectangles, respectively. A total of 3600 spectra were recorded, i.e., 200 spectra for each one of the 18 different voltage amplitudes. Distinct clusters of data points corresponding to the various voltage amplitudes are formed along the PC1 axis (describing 99.7% of the original data variance), whereas along the PC2 axis (0.1% of the variance), a clear differentiation based on the type of applied voltage to the SDBD is achieved. These results support the feasibility of creating an unsupervised predictive model based on the different PCs to successfully identify discharge operating conditions that is otherwise unfeasible through a traditional analysis of the data provided by conventional diagnostics.

**Fig.5b** shows the PCA loadings, i.e., the coefficients of the linear combination of the original variables from which the PCs were constructed. The loadings of PC1 show a negative correlation of the overall spectral fingerprint with the PC1 values. Considering this and by looking at **Fig.5a**, in which the clusters are positioned along PC1 with increasing voltage level, it can be confirmed that the OES intensity and, thus, the density of excited states of emissive species (see below), can be considered as a major indicator of a change in the voltage amplitude. As a matter of fact, increasing the amplitude of the applied voltage to the powered electrode leads to an increased power deposition into the plasma. The external electric field applied to the electrodes will also be larger. This is expected to enhance the electron density/temperature and enable a more efficient energy transfer from electrons to gaseous species (mainly $N_2$ and $O_2$ in our case). With increasing voltage amplitude, more energetic electrons will be made available for triggering key reaction pathways for discharge sustainment such as direct molecular excitations and ionizations. This means that the population density of key gaseous species into different excited (e.g., $N_2$(A), $N_2$(B), $N_2$(C)) and ionised (e.g., $N_2^+$(X), $N_2^+$(B), …) emissive states will be modified at higher voltages which will be reflected to the intensities of emissive species in the OES spectrum. To better illustrate this, the general expressions for the densities of $N_2$(C) and $N_2^+$(B) excited states of the dominant radiative transitions in **Fig.4c,d** are exemplarily shown [70]:

$$[N_2(C)] = \frac{I_{SPS(v'\rightarrow v'')} \times \lambda_{SPS(v'\rightarrow v'')}}{C_{SPS(v'\rightarrow v'')} \times h \times c \times A_{SPS(v'\rightarrow v'')}}$$

$$[N_2^+(B)] = \frac{I_{FNS(v'\rightarrow v'')} \times \lambda_{FNS(v'\rightarrow v'')}}{C_{FNS(v'\rightarrow v'')} \times h \times c \times A_{FNS(v'\rightarrow v'')}}$$

where, for a given transition of a species $X$ from a vibrational state $v'$ of an upper electronic state to a vibrational state $v''$ of a lower electronic state at a wavelength $\lambda_X$, $I_X$ is the emission intensity, $C_X$ a constant which depends on the spectral response of the system at $\lambda_X$, $[X]$ the density of the upper state (excited or ionised), and $A_X$ the Einstein coefficient of the specific transition. From these expressions, the density of an upper state is proportional to the corresponding emission intensity recorded. Thus, an increased intensity of a species observed in the OES spectrum of the SDBD with increasing voltage (**Figs.4c,d**) is directly associated with an increased density of the corresponding excited state.

Furthermore, loadings of PC2 serve to interpret the differentiation between the two types of applied voltage. As can be seen from the figure, this distinction is related to the $N_2$ bands' structures, which, even though appear to be similar to the naked eye (**Fig.4e**), still are sufficiently different for the PCA to discriminate them. Provided that the collisional quenching by air species is the same for both cases, the contribution of different reaction mechanisms to the population of excited/ionised states depends on the peculiar features of each voltage waveform applied to the SDBD ($\Delta V/\Delta t$, polarity, amplitude). Therefore, the proposed PCA–driven analysis of SDBD emission spectra seems capable of detecting imperceptible changes in the production mechanisms of excited states of key emissive species without the need of studying complex production/depopulation processes.

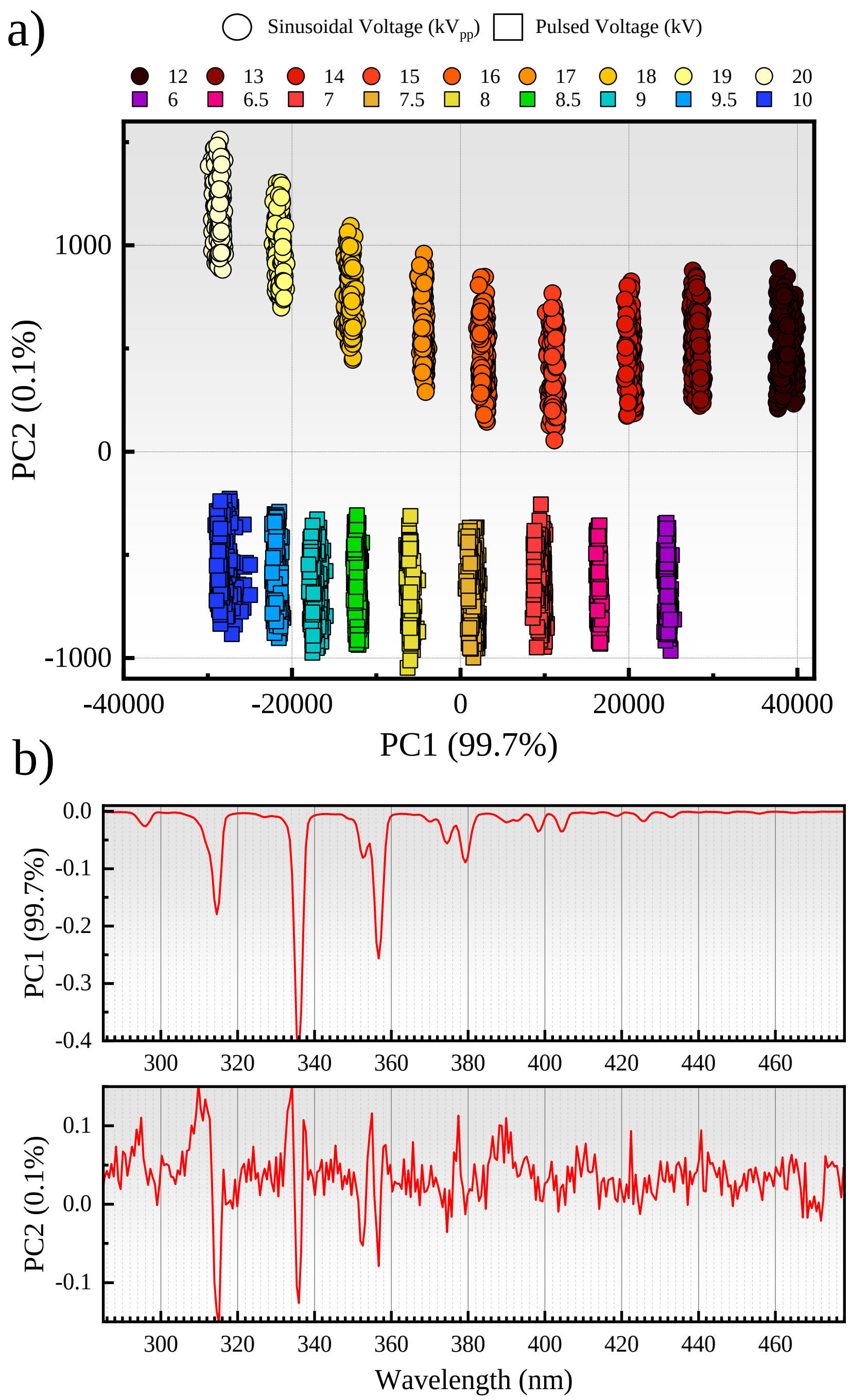

***Fig. 5.*** *a) PCA plot of SDBD emission spectra (3600 in total); SDBD orientation: 0° angle (**Fig.1a**). b) Loadings of the first two PCs against the wavelength.*

Here it could be argued that the pre-processing of the spectroscopic data may affect the PCA results. Thus, to consolidate the validity of the present methodology, we considered the worst-case scenario of data manipulation, referring to an individual normalization of each spectral profile (0–1 scale), and assessed its effect on the results of **Fig.5**. **Fig.6a,b** show the PCA-processed data after normalizing the experimental spectra recorded. In the PCA plot of **Fig.6a** the clusters have a noticeable overlap between them in contrast with **Fig.5a**. This is however expected because unnormalized data have a large contribution to PC1 with their overall spectral intensity values (PC1 loadings plot of **Fig.5b**), which in this case is eliminated through the intensity normalization process. Despite the rather large noise levels in the loadings plots of **Fg.6b**, several small peaks are still detectable at the same wavelengths as in **Fig.5b**. Thus, by inspecting the PC1 and PC2 loadings in **Fig.6b**, it can be stated that the features that most contribute to the cluster formations are hidden in the spectral envelopes of $N_2$(SPS) and $N_2^+$(FNS).

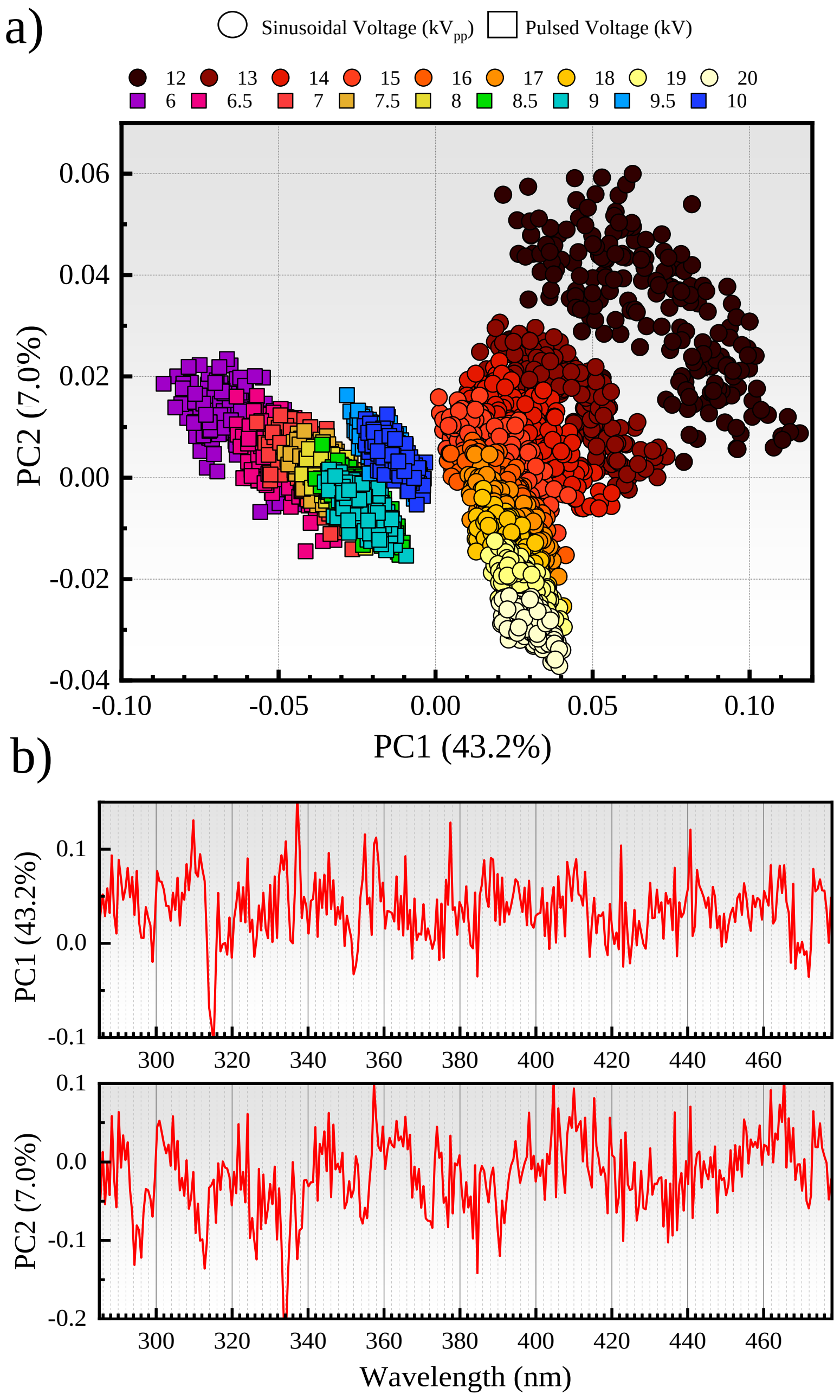

***Fig. 6.*** *a) PCA plot of normalised spectra of* ***Fig.5*** *and b) loadings of the first two PCs against the wavelength for the normalised spectra.*

To support the findings from the PCA–assisted treatments of OES spectra and to demonstrate the versatility and robustness of this method, a similar PCA transformation was performed to a series of microscope images (**Fig.7a,b** and **Table 1**). In total, 9560 images were recorded, i.e., 1195 images for each one of the 8 different voltage amplitudes (**Fig.7a**). To be noted that no data pre-processing was performed for these analyses. In this case, a clear discrimination is performed based on the captured discharge visible patterns in the CCD images. Indeed, **Fig.7a** illustrates distinct clusters of data points indicative of the different voltage types and amplitudes. Differentiation between the applied voltage amplitudes is evident along PC1 while a separation of the two voltage modes is seen along PC2. **Fig.7b** shows the corresponding loadings obtained from PCA (inverted images). It reveals that the main dissimilarities between the two voltage types are due to the formation of well–separated emission channels in the case of the pulsed voltage. These are not clearly observable for the AC voltage which exhibits more stochastic emission patterns (see also **Fig.4a,b**). Therefore, the PCA algorithm is able to identify small changes in the discharge emission intensity and pattern and achieve a fast classification, which would not be easily achievable through a manual treatment of this large quantity of images.

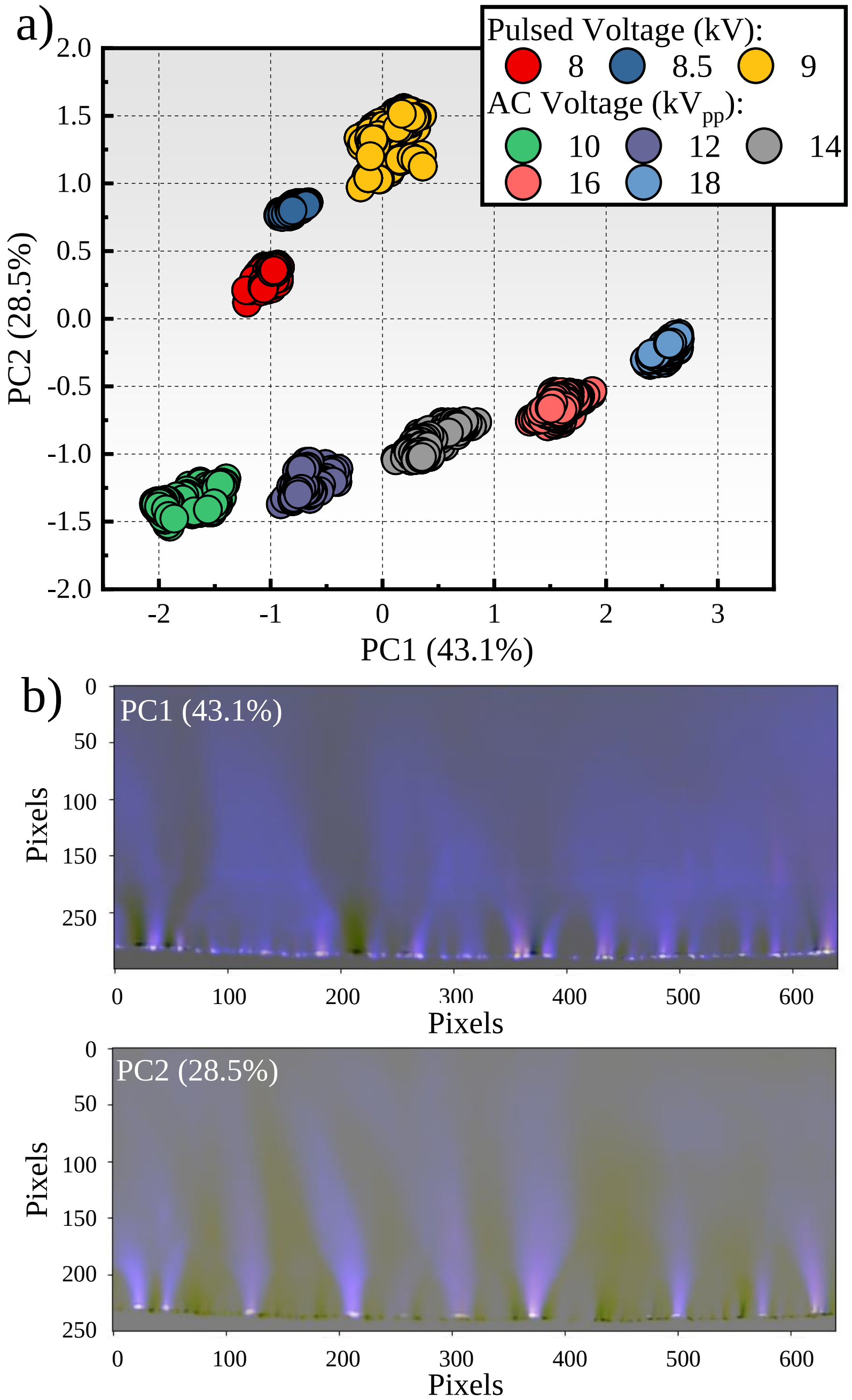

***Fig. 7.*** *a) PCA plot of SDBD emission captured by CCD images, b) loadings of the first two PCs of CCD images (9560 in total).*

**C. Voltage amplitude prediction at 0° angle using trained MLP at 0° angle.**

It was demonstrated that PCA can achieve clustering/discrimination of an ensemble of data corresponding to various voltage types and amplitudes applied to the SDBD, see **Fig.5,6,7**. However, when it is informed by only one group of spectra/images which explicitly refer to the same voltage type and amplitude, it is incapable of predicting the actual voltage amplitude applied to the SDBD. To achieve such a prediction, a coupled approach should be used by combining PCA with other ML architectures. In this work, as a proof of concept, 3600 experimental PCA–pre-processed OES spectra recorded at the reference position (**Fig.5**) were used to train MLP models and predict the actual voltage amplitude applied to the SDBD (**Tab.1**). The decision of using PCA–pre-processed OES spectra was performed mainly because only a few PCs describe most of the data's variance and, in this sense, the input variables can be significantly reduced from 423 (i.e., number of pixels in the unprocessed spectra) to only a few variables. Thus, the raw PCA–pre-processed spectra (i.e., without normalization) were fed to the network for subsequent analysis.

To define the right MLP parameters for an optimised prediction, first a hyperparameter optimisation was performed via the grid search method by testing various parameters. The best results were obtained by using 100 PCs as input and the MLP has an architecture of four hidden layers with 100, 100, 100 and 50 neurons in the first, second, third and fourth layer, respectively. The weight optimiser used was the Adam algorithm while *tanh* was the activation function.

The robustness of the predictive model was tested by iteratively removing one of the 18 voltage amplitudes (i.e., 9 for AC and 9 for pulsed) in each iteration to be used as a test set, while training with the rest 17, in order to ensure that the model is verified on unknown data corresponding to a specific voltage amplitude. This is shown in **Fig.8a**, where the 18 classes are found at the top being represented with different colours. An indicative example of the result of this procedure is shown in **Fig.8b**, where the predicted voltage amplitudes are plotted against the actual ones. By fitting the data with a linear function (red line in **Fig.8b**) the coefficient of determination ($R^2$) was found to be ≈0.99, indicating a very good fitting of the data. Therefore, by processing with the PCA the spectroscopic data recorded at the reference position and using the PCA outputs for training the MLP model, an accurate prediction of the actual voltage amplitude applied to the SDBD was achieved. This also shows a complementarity between the two ML methods used in this study.

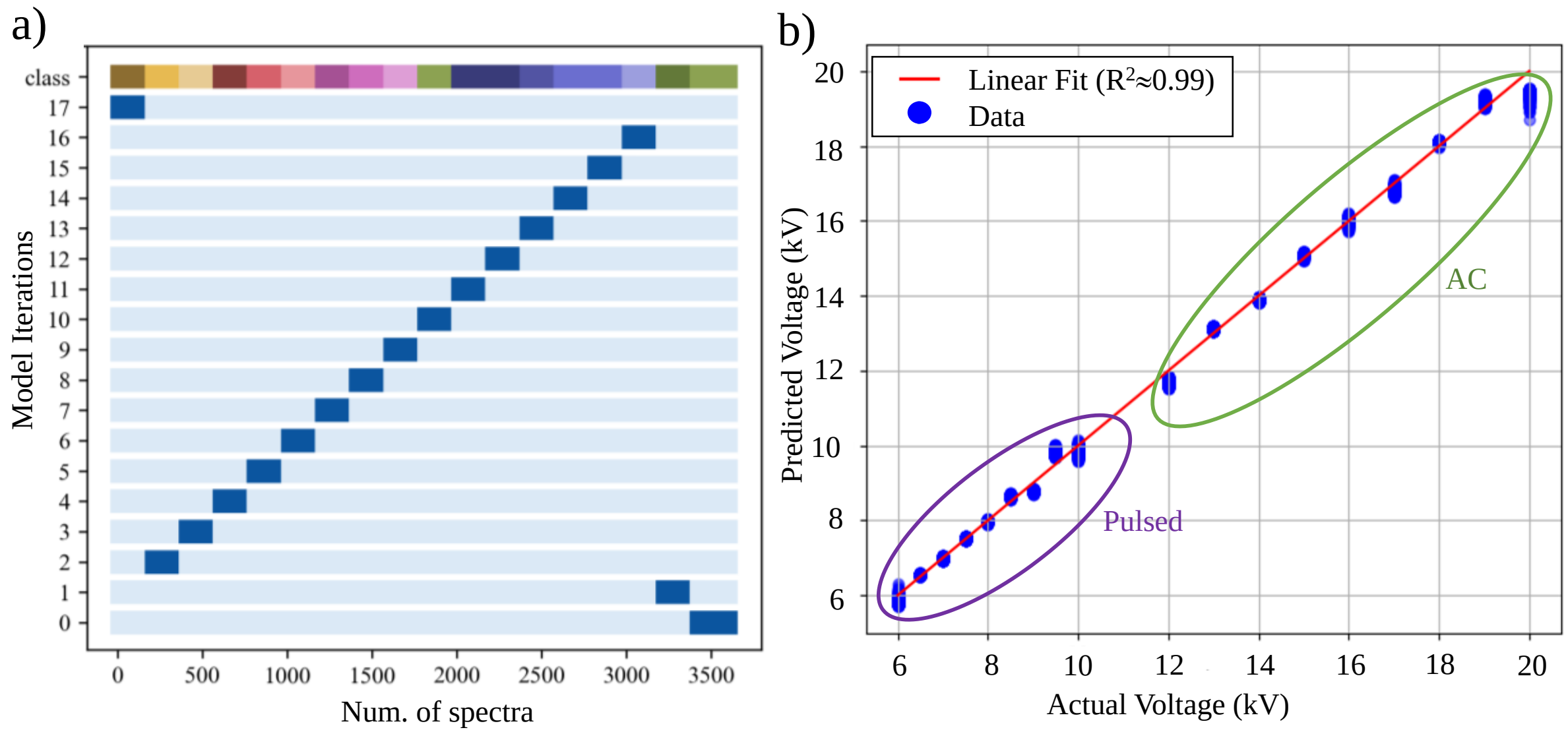

***Fig. 8.*** *a) Visualization of the MLP validation procedure where predictive models are tested by iteratively removing one of the voltage values to be used as a test set and training with the rest. b) Neural network–predicted voltage amplitudes plotted against the actual ones (each point corresponds to the average prediction of the 200 spectra recorded at 0° angle).*

**D. PCA and MLP for assessing the uniformity of the SDBD emission.**

For a more comprehensive examination of the discharge behaviour under varying amplitudes of the two waveforms of the applied voltage, additional experiments were conducted to determine whether its emission pattern around the quartz surface is uniform or not. To assess this, PCA was employed to visualize whether the SDBD spectra are clustered around specific points based on the applied voltage (see **Fig. 5a**), while the MLP constructed in the preceding section (see **Fig.8b**) was utilized to evaluate the accuracy of predicting the actual voltage amplitude applied to the SDBD. In this context, the uniformity of the emission would be apparent in cases where accurate predictions are made.

To this end, a total of 14400 OES spectra were recorded around the cylindrical surface (**Tab.1**). This was achieved by rotating the reactor by 90°, 180° and 270° with respect to its initial orientation (0° angle; **Fig.1**). The evaluation was performed by PCA to transform and project the obtained spectra at the initially–created PCA space (see **Fig.5a**, i.e., data obtained at 0°). **Fig.9** is the same as **Fig.5a** except that the spectra recorded at 90° are projected into the PCA space occupied by the spectra recorded at 0°. In the case of the AC mode (down–pointing triangles), the projected spectra along PC1 are placed very close to their corresponding clusters (i.e., those recorded at 0°), but there is a more pronounced displacement along PC2. For the pulsed mode (asterisks), for V≥7 kV the spectra at 90° are relatively well placed within the clusters of the spectra referring to 0°, showing a good level of emission uniformity, but for lower voltage amplitudes, the projected spectra are clearly misplaced. Based on the loadings (**Fig.5b**) it can be deduced that the overall emission intensity at lower voltage values differs significantly between 0° and 90°, since the projected points are clustered farther away from the cluster of points corresponding to 0°. Therefore, the emission pattern does not exhibit the same level of uniformity between these two SDBD regions. Similar observations can be made when projecting the spectra recorded at 180° and 270°, which are not included to reduce the complexity of the plot as much as possible.

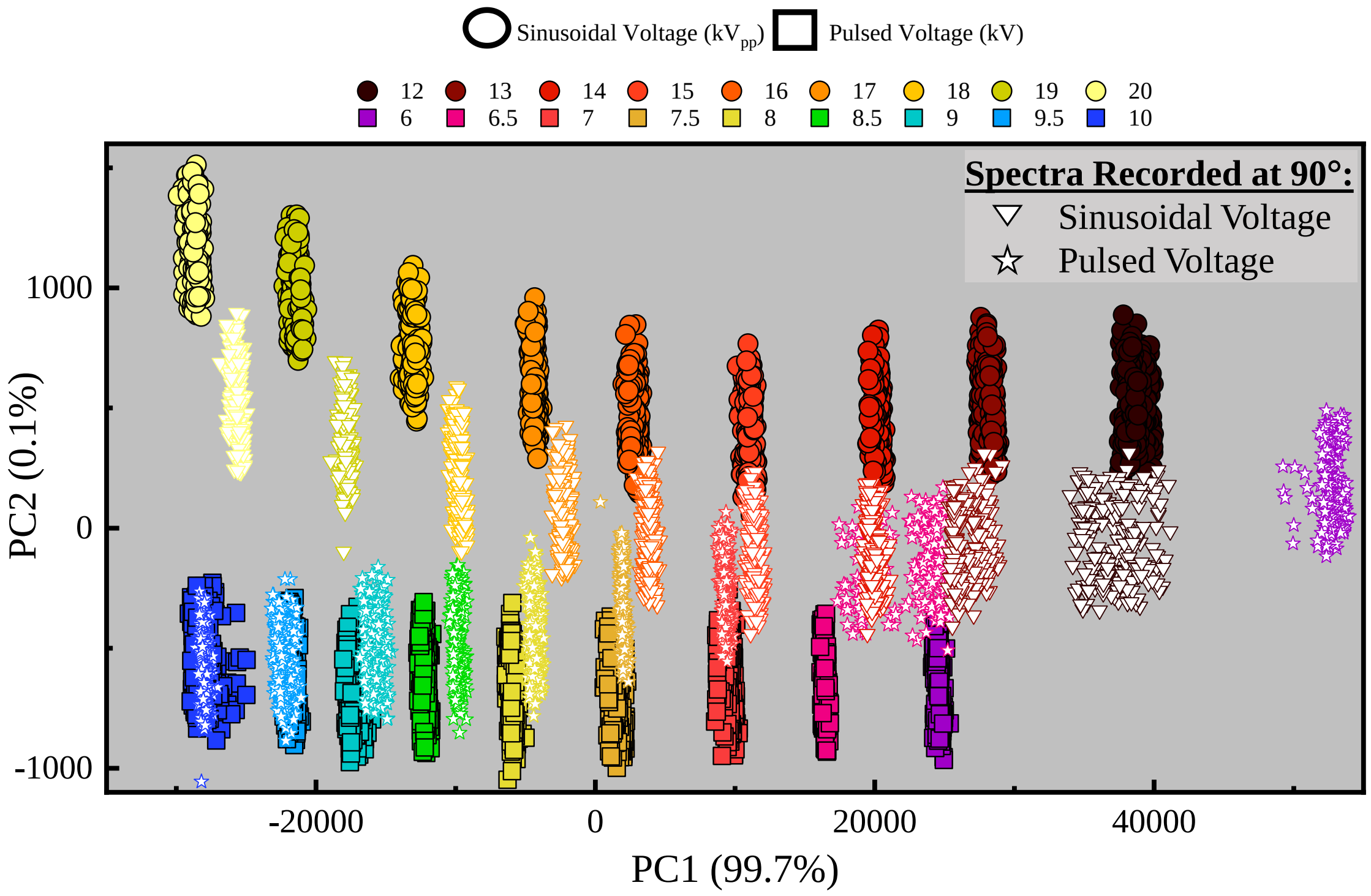

***Fig.9.*** *PCA plot of SDBD emission spectra at 0° (circles and rectangles; same with* ***Fig.5a****) along with the projection of spectra recorded at 90°.*

The MLP predictions of the voltage amplitudes from the OES spectra recorded at additional angles around the reactor are given in **Fig.10**. Each box corresponds to the predictions of 200 spectra for each angle and voltage type/amplitude, showing its corresponding statistics (see legend). Despite the accurate results achieved at 0° where MLP was trained by PCA–treated data at the same angle (**Fig.8**), in **Fig.10** the model is intentionally not trained for the additional angles. This makes it less accurate in predicting the actual voltage amplitude corresponding to additional angles under both modes. Specifically, for the pulsed mode and V≥8 kV, a relatively good level of prediction is achieved, while for V<8 kV the prediction error progressively increases. For the AC mode, larger variations of the predictions can be noticed. Despite its rather limited prediction power in these cases, the MLP model can still be used as a tool to efficiently detect differences in the discharge pattern between different angles around the reactor for a given voltage waveform and amplitude. For instance, the results shown in **Fig.10** for the lowest amplitudes of the pulsed mode (V<8 kV) indicate a strongly non uniform discharge emission pattern around the reactor. Indeed, in these cases, especially at 6 and 6.5 kV, some dark zones (i.e., absence of discharge emission channels) were clearly observed around the SDBD, thus affecting the discharge uniformity.

Here it is emphasized that the MLP is a supervised learning algorithm. Although it has the advantage to learn non–linear patterns in real–time, its validation accuracy may be affected due to random weight initializations, it is sensitive to feature scaling, etc. To improve its prediction accuracy and, thus, its reliable use for predicting the voltage amplitude, it would be necessary to include some spectra recorded at different angles within the training which is done in section **IIIE**. Such training of the MLP model could make it a valuable tool for applications aiming at a real–time monitoring of discharge features [16].

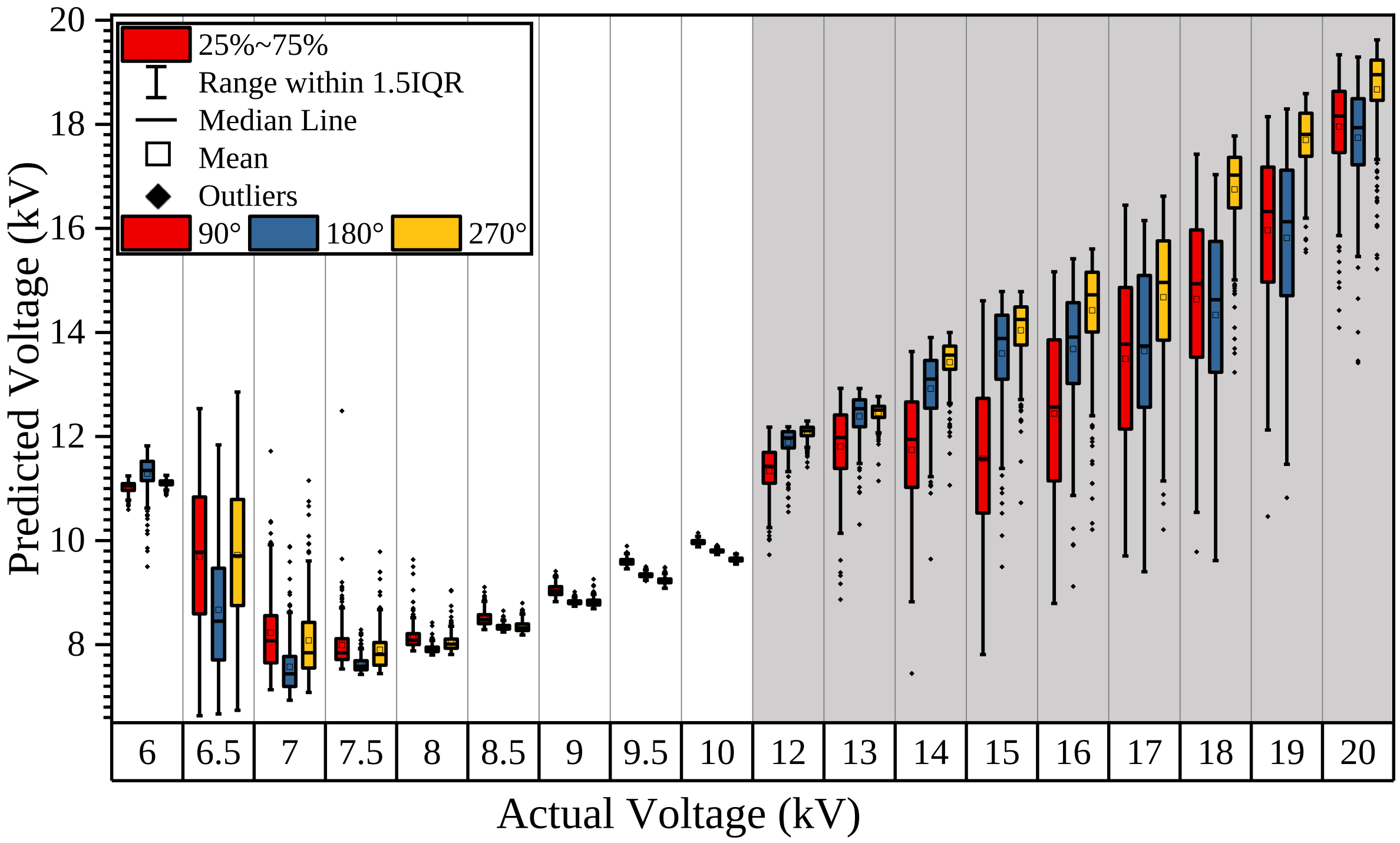

***Fig.10.*** *Neural network prediction of the actual voltage amplitudes at different angles (i.e., 90°, 180°, 270°) of the reactor for pulsed (6–10 kV; white columns) and AC (12–20 kV$_{pp}$; gray columns) cases. The MLP is only trained with data recorded at 0°.*

**E. Intermediate voltage amplitude prediction at 0° angle using trained MLP at all angles.**

In the previous section it was shown that the prediction accuracy of the actual voltage amplitude using MLP strongly depends on its training. Therefore, we aimed to improve its prediction error. To achieve this, the variation of the emission spectra around the reactor was considered during the building process of the model. For this purpose, the neural network architecture presented in Section **IIIC**, was trained using a total of 14400 spectra (i.e., 3600 spectra recorded at each of the four different angles around the reactor). In order to assess the prediction accuracy of the MLP, a total of 3200 spectra were recorded at 0°, using different voltage values than those used for the training of the model. Specifically, these spectra were recorded for 16 voltage levels (i.e., 8 for the pulsed and 8 for the AC power supply), obtaining 200 spectra for each level.

In **Fig.11** the predictions of the voltage values based on the aforementioned spectra are shown for the pulsed (**Fig.11a**) and the AC (**Fig.11b**) power supplies. The *x*-axes correspond to the actual voltage values and the *y*-axes to the predicted values from the MLP. As can be seen, the predicted voltage amplitudes, which were not included in the training of the algorithm, are precise with small deviations from the actual voltage amplitudes applied to the SDBD. Thus, it can be confirmed that MLP can be a powerful tool for predicting actual discharge operating conditions upon appropriate training.

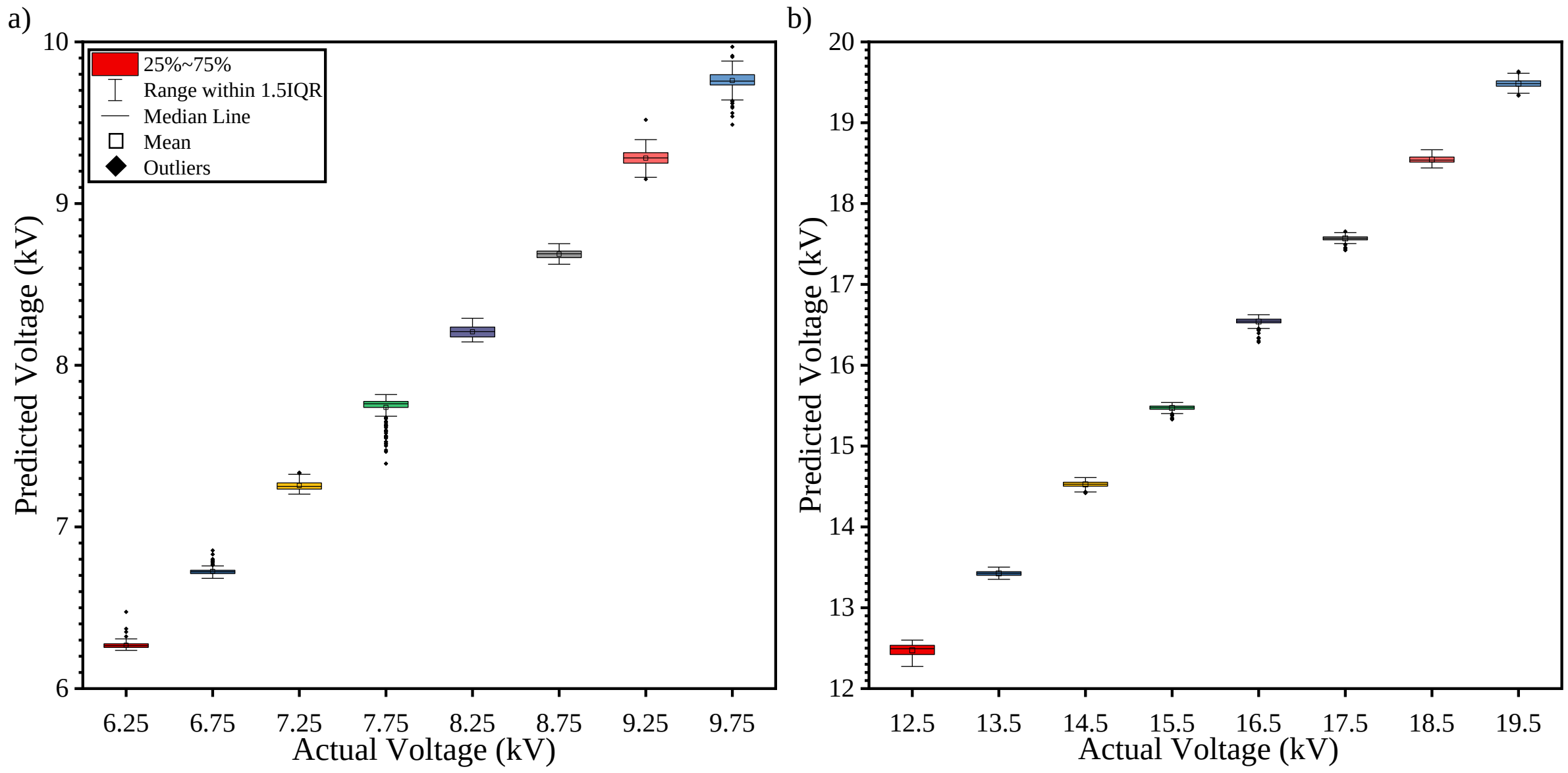

***Fig.11.*** *Neural network prediction of voltage amplitudes obtained at 0° for a) pulsed and b) AC voltage waveforms.*

## IV. CONCLUSIONS

This study demonstrates the strong complementarity of simple optical diagnostics (such as time–integrated OES and CCD imaging) with unsupervised (PCA) and supervised (MLP) machine learning algorithms for a more comprehensive characterisation of a cylindrical atmospheric pressure SDBD. Through this coupled approach, it is showcased that the discharge's operating conditions (applied voltage waveform and amplitude) can be precisely differentiated and predicted by identifying hidden correlations/differences within large amounts of spectroscopic/imaging data recorded. The differentiation of the voltage waveform and amplitude applied to the SDBD is achieved using PCA which transforms the discharge's emission spectra and CCD images into principal components (PC) and projects them in a two-dimensional PC space. The prediction of the voltage amplitude cannot be achieved using the PCA, and it is done using the MLP artificial neural network, which is trained using PCA–preprocessed data. Furthermore, both the PCA and the MLP allow assessing the level of uniformity of the discharge's emission pattern at distinct regions around the reactor. These findings underscore the capacity of ML in achieving unique classifications and accurate predictions of various operating conditions of SDBDs towards a real–time control and monitoring of global plasma parameters for different applications. The present plasma device and proposed data–driven methodologies may be particularly suitable for aerodynamic flow control applications.

## Acknowledgements

This work was funded by the Labex SEAM project (ANR–10–LABX–0096; ANR–18–IDEX–0001), the ANR ULTRAMAP project (ANR–22–CE51–0027), the IDF regional project SESAME DIAGPLAS, and partly by the DFG project MAID (number 466331904).